\documentclass[sn-nature]{sn-jnl}% Style for submissions to Nature Portfolio journals
\usepackage{graphicx}%
\usepackage{multirow}%
\usepackage{amsmath,amssymb,amsfonts}%
\usepackage{amsthm}%
\usepackage{mathrsfs}%
\usepackage[title]{appendix}%
\usepackage{xcolor}%
\usepackage{textcomp}%
\usepackage{manyfoot}%
\usepackage{booktabs}%
\usepackage{algorithm}%
\usepackage{algorithmicx}%
\usepackage{algpseudocode}%
\usepackage{listings}%
\usepackage{svg}
\usepackage{gensymb}
\usepackage{lmodern}
\usepackage{subcaption} 
\begin{document}

\title[Article Title]{Emerging clean technologies: policy-driven cost reductions, implications and perspectives}

\author*[1]{\fnm{Mohamed} \sur{Atouife}}\email{m.atouife@princeton.edu}

\author*[2]{\fnm{Jesse} \sur{Jenkins}}\email{jessejenkins@princeton.edu}
\affil[1]{\orgdiv{Mechanical and Aerospace Engineering}, \orgname{Princeton University}, \city{Princeton}, \state{New Jersey}, \postcode{08540}, \country{United States}}
\affil[2]{\orgdiv{Mechanical and Aerospace Engineering}, \orgname{Princeton University}, \city{Princeton}, \state{New Jersey}, \postcode{08540}, \country{United States}}
\maketitle


\section*{Context \& Scale}
Early stage policy support is critical for the demonstration, commercialization and diffusion of nascent technologies. Policy-induced deployment unlocks cost reductions through learning-by-doing, driving further deployment and greater cost reductions, thus enabling a positive feedback loop. Eventually, markets and supply chains are established, the technology matures and government support is phased out.
Electrolysis and direct air capture, two emerging technologies crucial to reaching net-zero targets, have received significant policy backing and are at the early phases of this technological cycle. In this context, we project policy-driven cost declines of these nascent technologies through 2030, assess their impact on the cost of electrolytic hydrogen production, carbon removal from the atmosphere, synthetic kerosene production from hydrogen and CO$_{\text{2}}$, and discuss policy implications and perspectives. 
\section*{Summary}
Hydrogen production from water electrolysis, direct air capture (DAC), and synthetic kerosene derived from hydrogen and CO$_{\text{2}}$ (`e-kerosene') are expected to play an important role in global decarbonization efforts. So far, the economics of these nascent technologies hamper their market diffusion. However, a wave of recent policy support in the United States, Europe, China, and elsewhere is anticipated to drive their commercial liftoff and bring their costs down. To this end, we evaluate the potential cost reductions driven by policy-induced scale-up of these emerging technologies through 2030 using an experience curves approach accounting for both local and global learning effects. We then analyze the consequences of projected cost declines on the competitiveness of these nascent technologies compared to conventional fossil alternatives, where applicable, and highlight some of the tradeoffs associated with their expansion. Our findings indicate that enacted policies could lead to substantial capital cost reductions for electrolyzers. Nevertheless, electrolytic hydrogen production at \$1-2/kg would still require some form of policy support. Given expected costs and experience curves, it is unlikely that liquid solvent DAC (L-DAC) scale-up will bring removal costs to stated targets of \$100/tCO$_{\text{2}}$, though a \$200/tCO$_{\text{2}}$ may eventually be within reach. We also underscore the importance of tackling methane leakage for natural gas-powered L-DAC: unmitigated leaks amplify net removal costs, exacerbate the investment requirements to reach targeted costs, and cast doubt on the technology's role in the clean energy transition. Lastly, despite reductions in electrolysis and L-DAC costs, e-kerosene remains considerably more expensive than fossil jet fuel. The economics of e-kerosene and the resources required for production raise questions about the fuel's ultimate viability as a decarbonization tool for aviation.

%%================================%%
%% Sample for structured abstract %%
%%================================%%

\keywords{Emerging technologies, technology policy, electrolysis, direct air capture, e-kerosene, experience curves, climate mitigation.}



\section*{Introduction}
\begin{figure*}%
    \centering
    \includegraphics[width=\columnwidth]{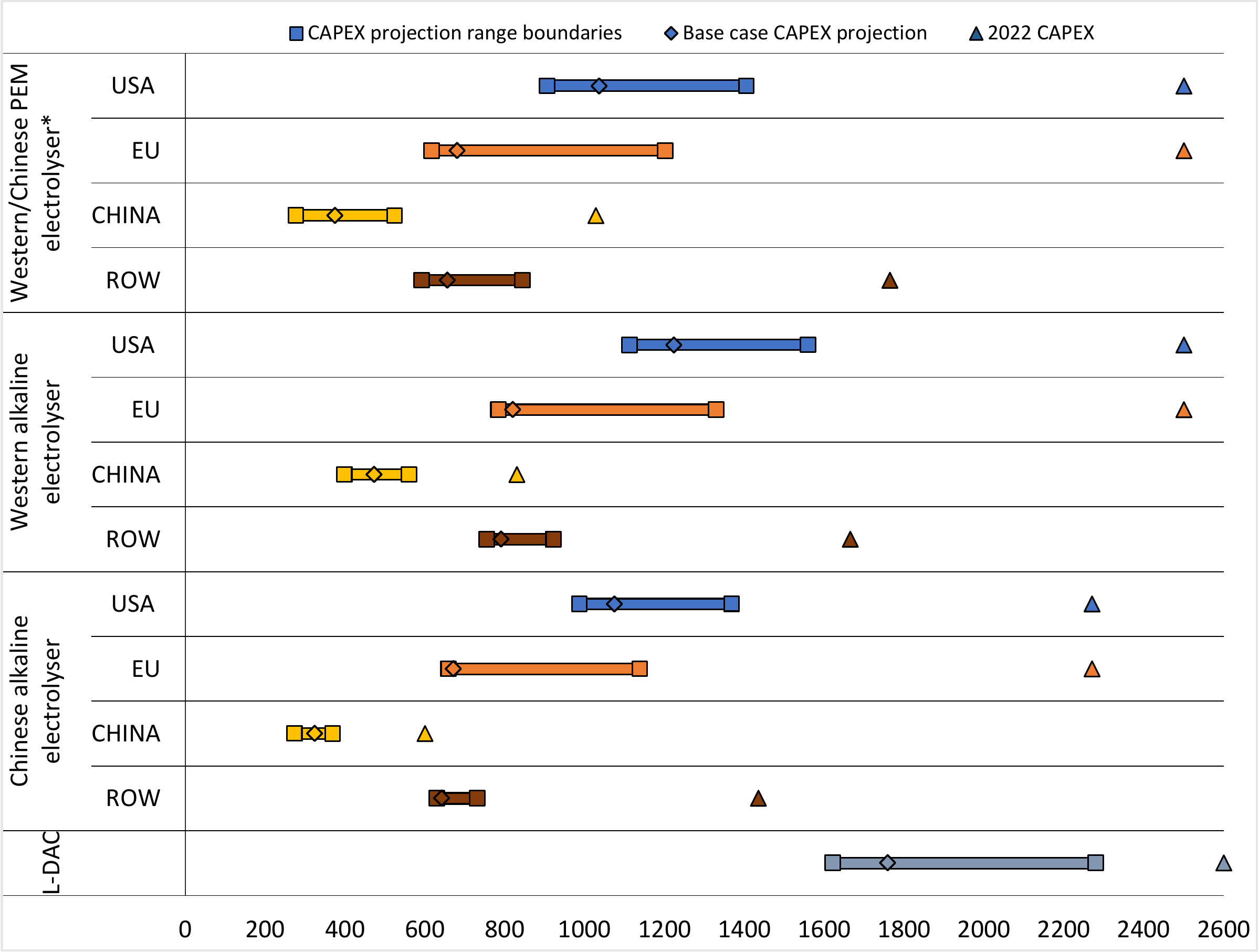}%
    \caption{\textbf{Current and projected costs of key emerging technologies in 2030 by region}. The range indicates sensitivity analysis for the assumed learning rates, and the deployed capacity in 2030. Electrolysis costs are in USD per kW and L-DAC costs are in USD per tCO$_2$/year capture capacity respectively. ROW stands for the rest of the world.\\ 
    *Assumes 50/50 shares of global deployment for Western/Chinese PEM, learning rates are identical; outcomes for alternative splits in global deployment are illustrated in the supplemental information and interactive dashboard. }
    \label{Fig 1:}
\end{figure*}

As emerging technologies, hydrogen production from water electrolysis, carbon dioxide removal via direct air capture (DAC), and synthetic kerosene derived from hydrogen and CO$_2$ (aka `e-kerosene') are at present in a similar state as solar photovoltaics and wind power were in their early days: expensive, uncompetitive technologies deployed only for niche applications. Strong policy support for early deployment allowed wind and solar to overcome barriers to market introduction, eventually diffuse, and experience rapid cost declines through learning by doing or `experience curves' \cite{de_la_tour_predicting_2013,schmidt_future_2017}. Experience curves describe deployment-induced cost reductions stemming from a wide range of factors: economies of scale in manufacturing, establishing and optimizing supply chains, incremental innovation in product design and manufacturing processes, and accumulating labor experience, to mention a few. Typically, the cost declines of mass-manufactured products, such as solar modules or lithium ion battery cells, result from innovations and economies of scale in manufacturing and are independent of where the products are eventually deployed. Their learning is contingent on a global market pull and therefore follows a global learning curve. On the other hand, the decline in costs of procurement and final installation is contingent on the accumulated experience accrued from local deployment of the technology and therefore typically follows a local learning curve \cite{huenteler_technology_2016,malhotra_accelerating_2020}.\\
Over the recent years, there has been a notable increase in the implementation of policies favoring electrolysis, DAC, and e-kerosene. Various regions, such as China and India, have outlined national hydrogen strategies with ambitious goals for renewable hydrogen production \cite{iea2023hydrogen}. In the United States, the Infrastructure and Jobs Act allocates multi-billion dollar grant funding to establish hubs for clean hydrogen and DAC \cite{doe2023hydrogenhub,energygovDAC}. Furthermore, the Inflation Reduction Act (IRA) incentivizes the production of clean hydrogen, synthetic fuels, and carbon dioxide capture from the atmosphere through generous tax incentives \cite{whitehousecleanenergy}. In Europe, the renewable energy directive mandates that by 2030, $42\%$ of hydrogen used in industry must be renewable, and $1.2\%$ of aviation fuel supply must be synthetic kerosene \cite{refueleuaviation,eu_renewable_energy_rules}. Additionally, the European hydrogen bank is set to support clean hydrogen projects through auction-based production subsidies \cite{euhydrogenbank}.\\ 
Now that a supportive policy environment is in place, water electrolysis, DAC, and low-carbon liquid fuels including e-kerosene are anticipated to see initial commercial deployment and scale-up and consequently are likely to see their cost decline due to policy-induced experience curves. In the remainder of the paper, we apply different experience curve models integrating both local and global learning to project the potential impact of current policies on the capital costs of electrolysis and DAC. We then explore the implications of these capital cost reductions on the levelized cost of electrolytic hydrogen, carbon removal from the atmosphere, and e-kerosene. A summary of current and projected costs is displayed in Fig. \ref{Fig 1:}. We also provide interactive dashboards that allow users to modify input assumptions (i.e. installed capacities, learning rates, etc.) and obtain updated projections for capital and levelized costs in real time \cite{hydrogendashboard,dacdashboard}. We encourage readers to explore these interactive dashboards in conjunction with this paper.
\section*{Results}
\subsection*{Water electrolysis}
Global hydrogen demand reached 95 million metric tonnes in 2022, $77\%$ of which comes from the refining and fertilizer industries \cite{iea2023hydrogen}. Conventional hydrogen production pathways based on natural gas reforming or coal gasification result in more than 900 million tons of CO$_{2}$ emissions each year \cite{iea2023hydrogen}, equivalent to about 2.5 percent of current global CO$_2$ emissions. Given the urgency of climate mitigation, it is imperative to decarbonize hydrogen production. Moreover, the fuel offers the potential to serve as a versatile zero-carbon energy carrier, reductant, and intermediate feedstock for synthetic hydrocarbon production to help decarbonize other sectors, including steel, aviation and shipping. This further underscores the need to develop and scale low-emission hydrogen production technologies. Electrolysis, a technology that splits water into oxygen and hydrogen using electricity, is emerging as a cleaner and more sustainable alternative to fossil fuel-based production. Recent support policies are anticipated to spur large deployment of electrolyzers worldwide \cite{bnef2030electrolyzer}. We investigate the potential impact of this policy-driven build-out on the capital costs of water electrolysis projects and hydrogen production.
\subsubsection*{Capital cost reductions}
\begin{figure*}[htbp]
    \centering
    \begin{subfigure}[b]{0.475\textwidth}
        \centering
        \includegraphics[width=\linewidth]{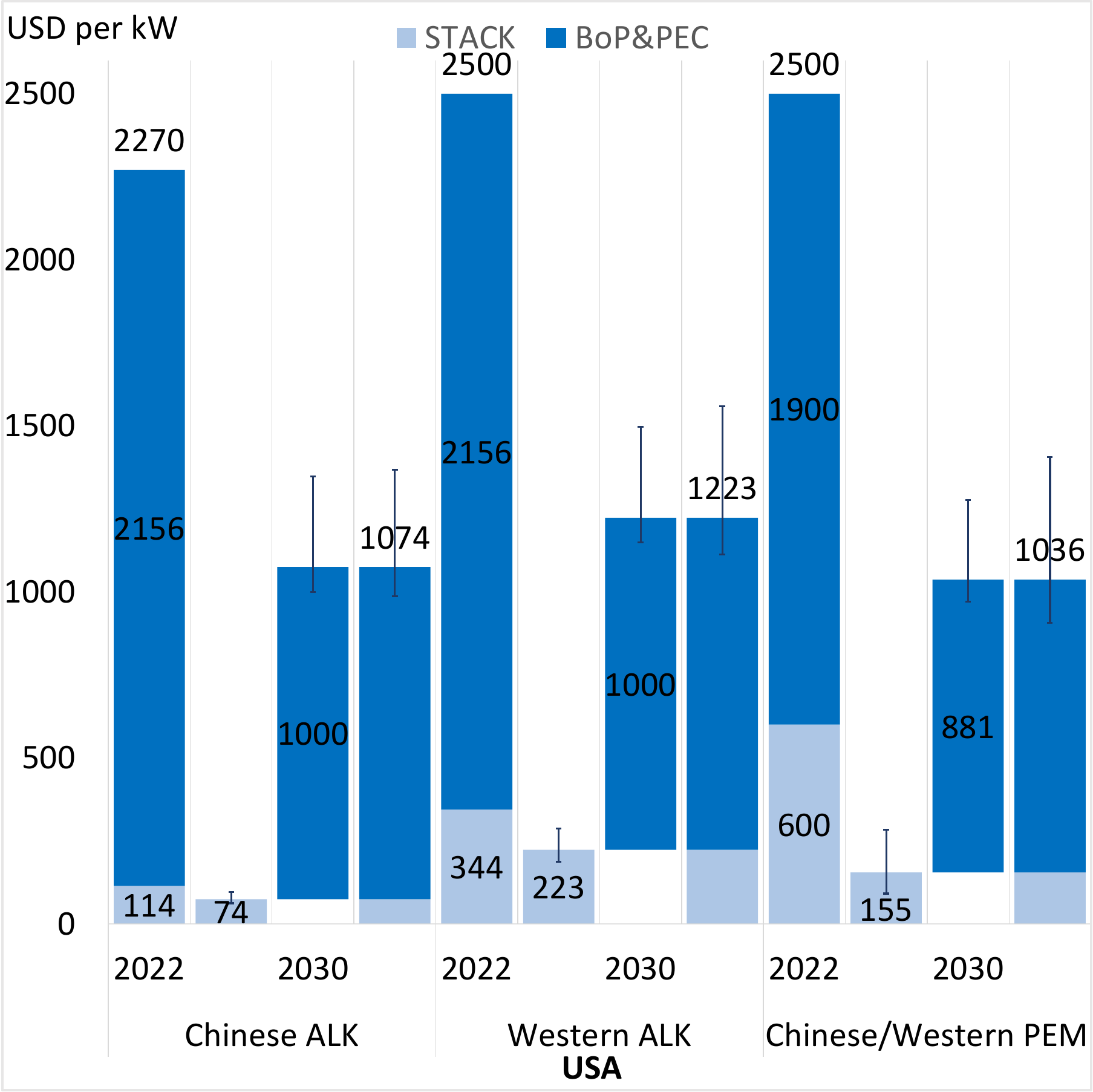}
    \end{subfigure}
    \begin{subfigure}[b]{0.475\textwidth}
        \centering
        \includegraphics[width=\linewidth]{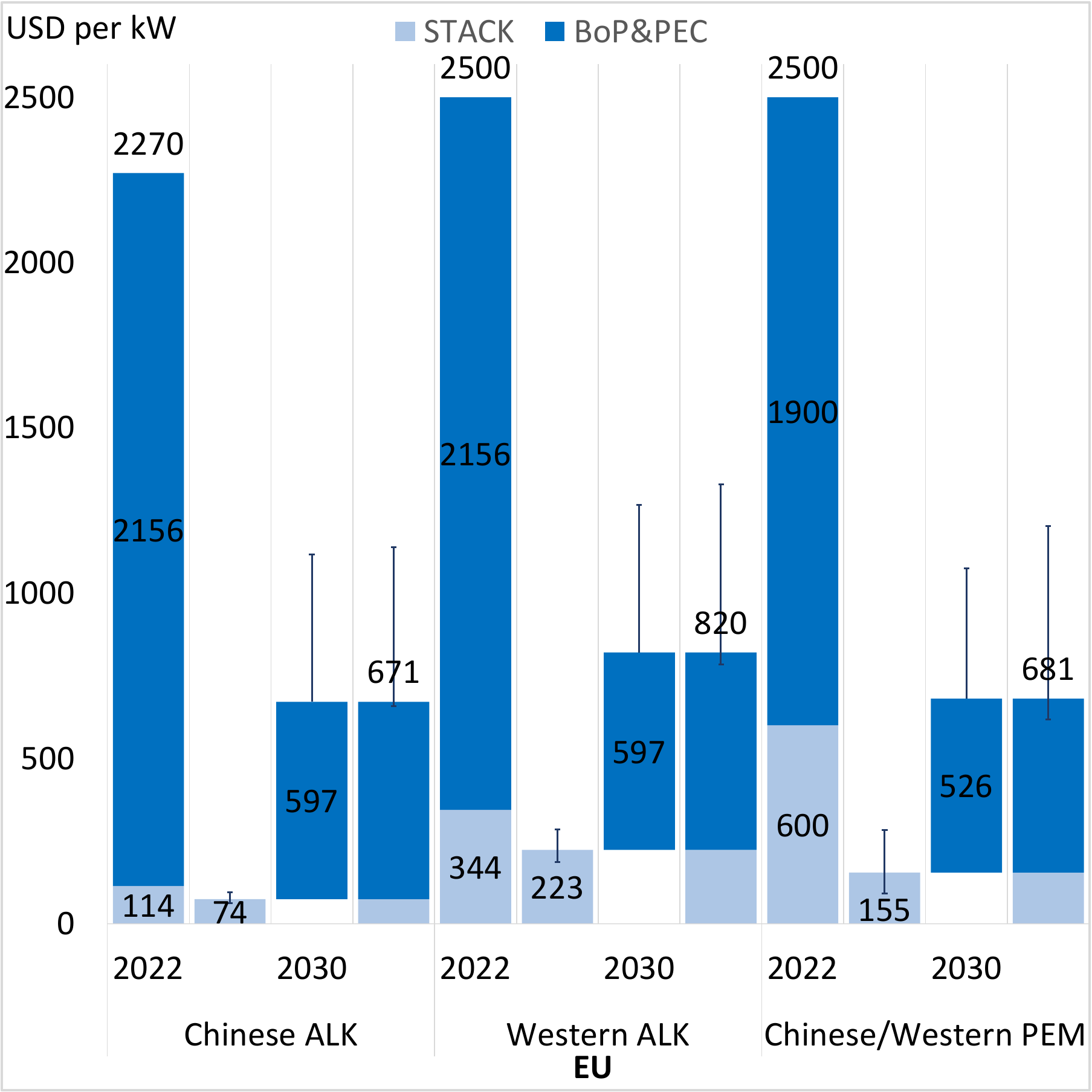}
    \end{subfigure}
    
    %\vspace{0.1cm} % Adjust vertical space between rows of figures
    
    \begin{subfigure}[b]{0.475\textwidth}
        \centering
        \includegraphics[width=\linewidth]{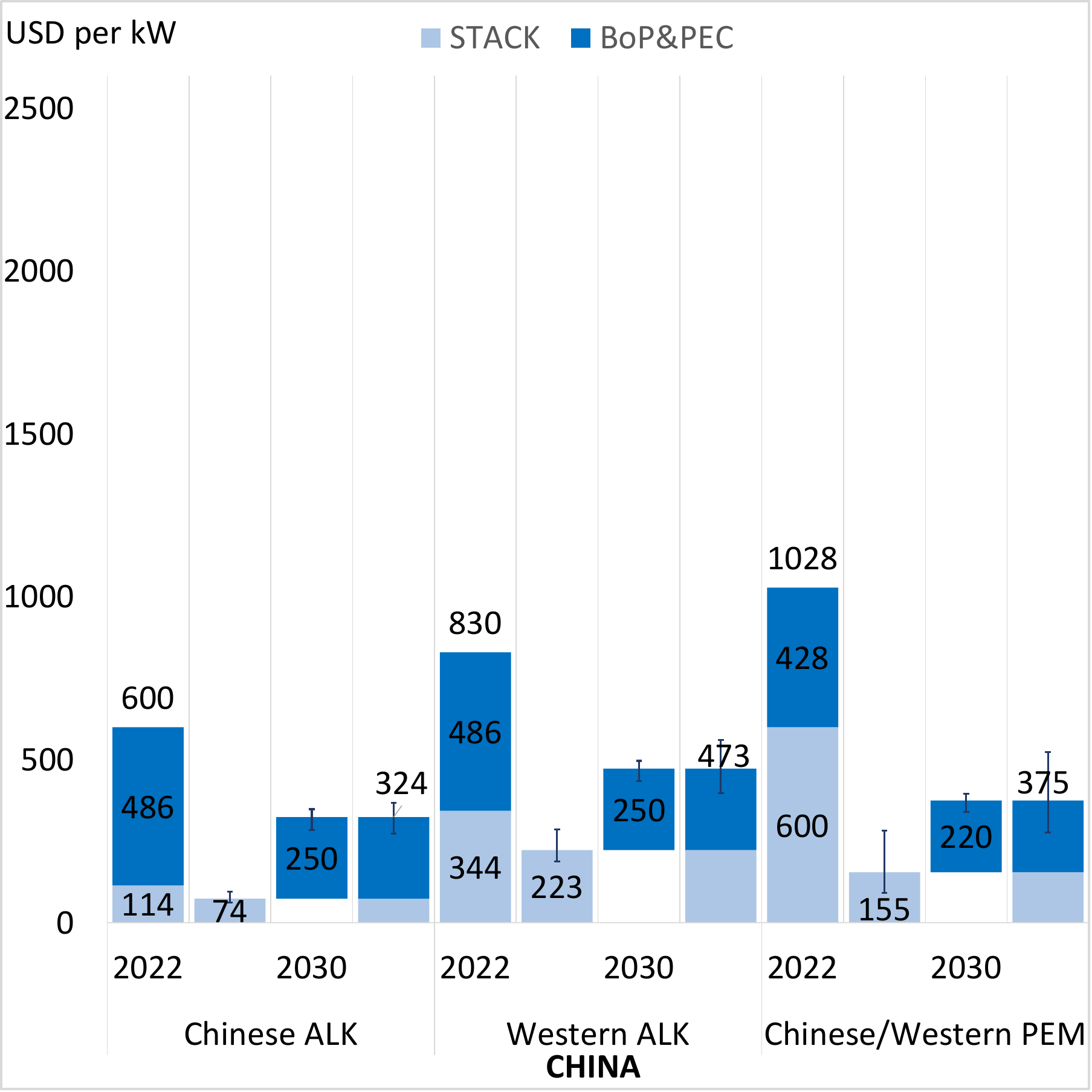}
    \end{subfigure}
    \begin{subfigure}[b]{0.475\textwidth}
        \centering
        \includegraphics[width=\linewidth]{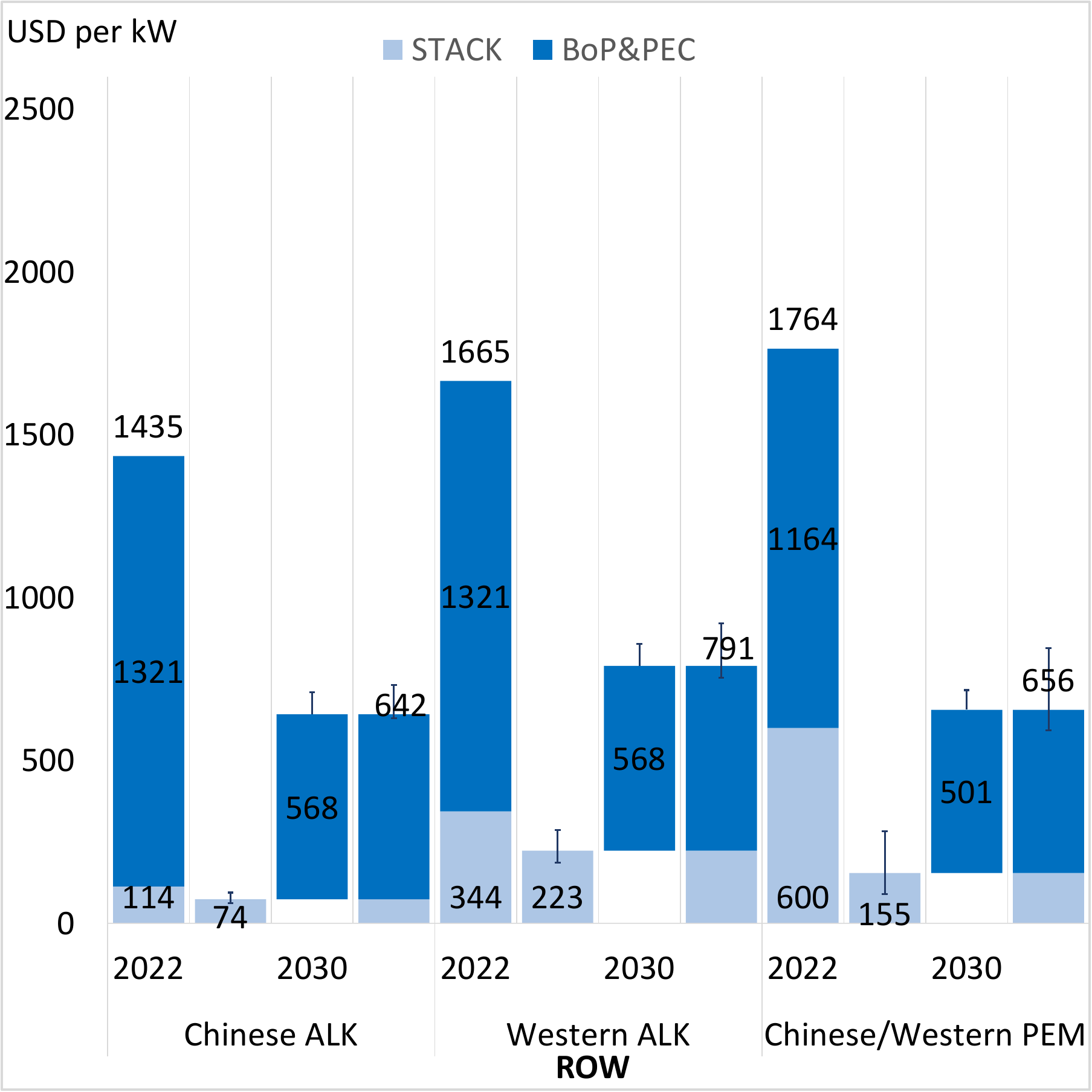}
    \end{subfigure}
    
    \caption{\textbf{Current and projected water electrolysis capital project costs by technology type and country/region.} The error bars depict sensitivity to the assumed learning rates as well as the installed electrolysis capacity in 2030.}
    \label{Fig 2:}
\end{figure*}

The capital costs of a water electrolysis project consist of three parts: the electrolyzer stack, the balance of plant (BoP), and engineering, procurement, and construction (EPC). The two principle electrolysis technologies, proton exchange membrane (PEM) and alkaline, not only differ in technological terms (use of separation membranes versus alkaline solution, ramp-up time, etc.) but also in terms of current maturity and costs. Additionally, reported alkaline stack costs currently differ substantially for Chinese and Western manufacturers (Fig. \ref{Fig 2:}). Since these mass-manufactured stack technologies compete for global deployment opportunities and have separate manufacturing supply chains, we assume that Western PEM, Chinese PEM, Western alkaline, and Chinese alkaline stack costs will each follow distinct global experience curves. Stated differently, the costs associated with a water electrolysis stack decrease as a function of the cumulative installed capacity of that specific stack technology on a global scale, and we assume Chinese and western supply chains develop largely independently. As for the BoP and PEC costs, Fig. \ref{Fig 2:} highlights regional differences stemming from local factors such as labor productivity, costs of project development, land costs, regulation, taxes, etc. For this reason, our experience curve model incorporates a local component: cost declines for the BoP and EPC for a specific region are contingent on the cumulative deployed electrolysis capacity in that region alone and progress at a region-specific learning rate (rather than a global rate as per stack costs). All stack technology costs are assumed to follow the learning rate empirically observed for alkaline electrolyzers, while the country-specific BoP and PEC costs follow country-specific BoP learning rates observed in the solar industry as a likely analog for electrolyzer deployment (see supporting information).\\ 
Looking ahead to 2030, we project a different technological landscape. Fig. \ref{Fig 2:} shows cost projections for the base case, where the deployment of each stack technology (alkaline versus PEM) is assumed to be evenly split between Chinese and Western technologies (see SI for results for varying market shares or explore the interactive dashboard). At the stack level, Chinese manufactured technologies continue to be cheaper than their Western alternatives, though their cost advantage diminishes. Furthermore, as a more nascent stack technology, PEM electrolyzers could undergo the greatest cost reductions, potentially becoming cheaper than western alkaline electrolyzers. Total installed electrolyzer system capital costs could decline by $41-74\%$ depending on the choice of stack technology and region.\\
%China remains the region with the lowest capital costs, though the cost advantage compared to other parts of the world significantly decreases. \\
What specific mechanisms are likely to lead to electrolyzer cost declines? The manufacturing of water electrolysis stacks currently occurs on a relatively small scale (with a few hundred megawatts deployed annually) and involves significant amounts of manual labor. Now, with a strong, policy-induced demand outlook, electrolyzer manufacturers are investing in gigawatt-scale factories that are likely to unlock substantial cost savings through economies of scale and automation \cite{iea2023hydrogen}. Additionally, the BoP and PEC components are likely to see costs fall as supply chains are established and optimized, more integrated systems are offered, larger-scale projects in the range of 100s of megawatts or gigawatts become more common, technology risk reduces, competition increases, and plant designs are standardized \cite{worley2023ambition}. On the other hand, prolonged periods of high-interest rates, uncertainty around clean hydrogen demand, and bottlenecks in rare earth material supplies (i.e. iridium and titanium for PEM) could slow the learning rate. Fig. \ref{Fig 2:} illustrates the impact of uncertainty in learning rates (and regional deployment), which can also be explored further in the interactive dashboard.

% \begin{figure*}%
%     \includegraphics[width=0.5\linewidth]{CAPEXUSA.pdf}%
%     \qquad
%     \includegraphics[width=0.5\linewidth]{CAPEXEU.pdf}%
%     \qquad
%     \includegraphics[width=0.5\linewidth]{CAPEXCHINA.pdf}%
%     \qquad
%     \includegraphics[width=0.5\linewidth]{CAPEXROW.pdf}%
%     \caption{\textbf{Current and projected water electrolysis capital project costs by technology type and country/region.} The error bars depict sensitivity to the assumed learning rates as well as the installed electrolysis capacity in 2030.}
%     \label{Fig 2:}
% \end{figure*}

\subsubsection*{Implications for electrolysis-based hydrogen production}
\begin{figure}[htbp]
    \centering
    
    \begin{subfigure}[b]{0.475\textwidth}
        \centering
        \includegraphics[width=\linewidth, height=4cm]{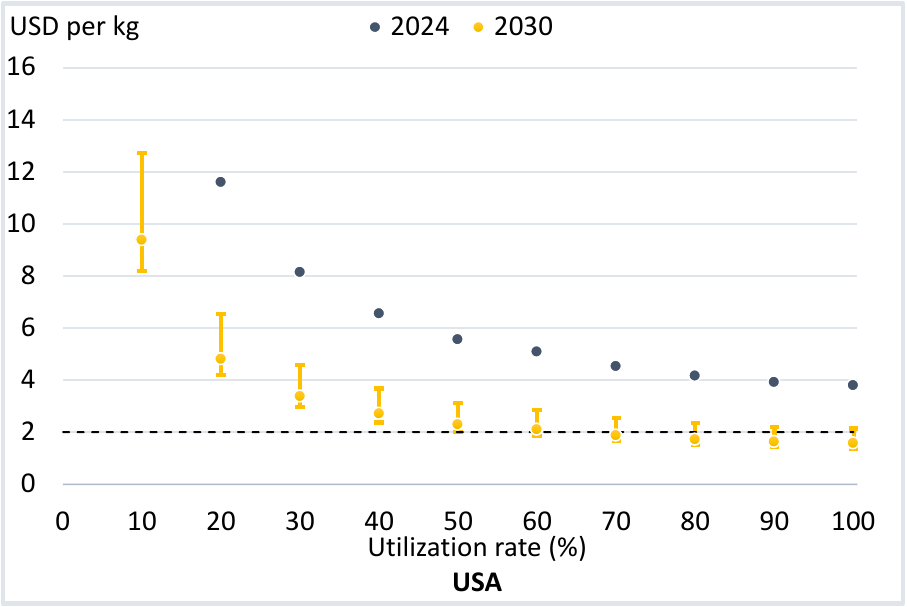}
    \end{subfigure}
    \hfill
    \begin{subfigure}[b]{0.475\textwidth}
        \centering
        \includegraphics[width=\linewidth, height=4cm]{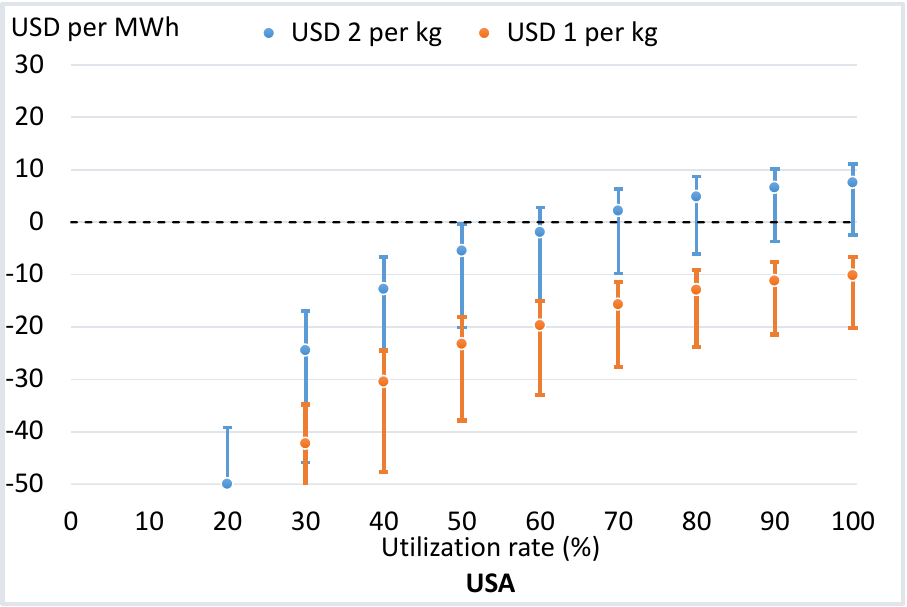}
    \end{subfigure}
    
    %\vspace{0.5cm} % Adjust vertical space between rows of figures
    
    \begin{subfigure}[b]{0.475\textwidth}
        \centering
        \includegraphics[width=\linewidth, height=4cm]{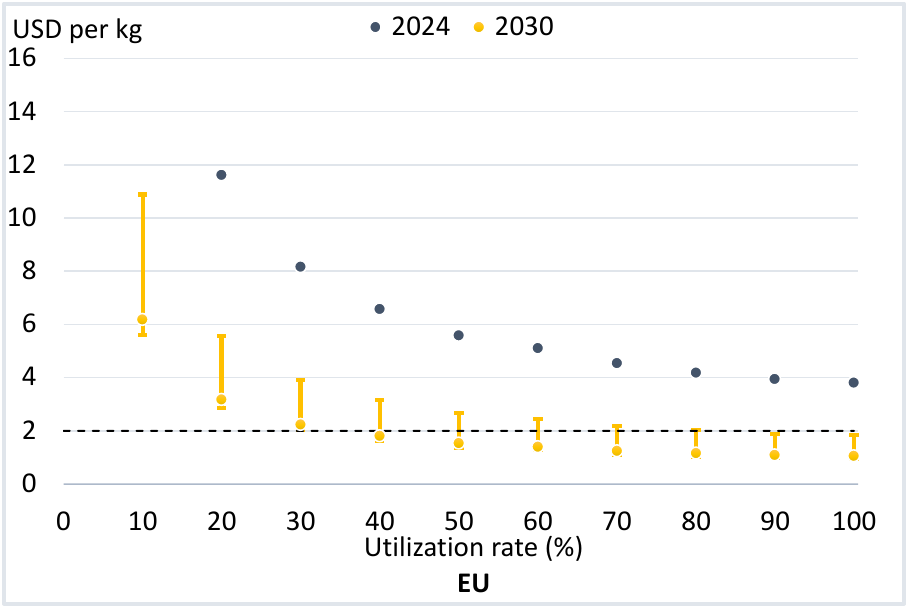}
    \end{subfigure}
    \hfill
    \begin{subfigure}[b]{0.475\textwidth}
        \centering
        \includegraphics[width=\linewidth, height=4cm]{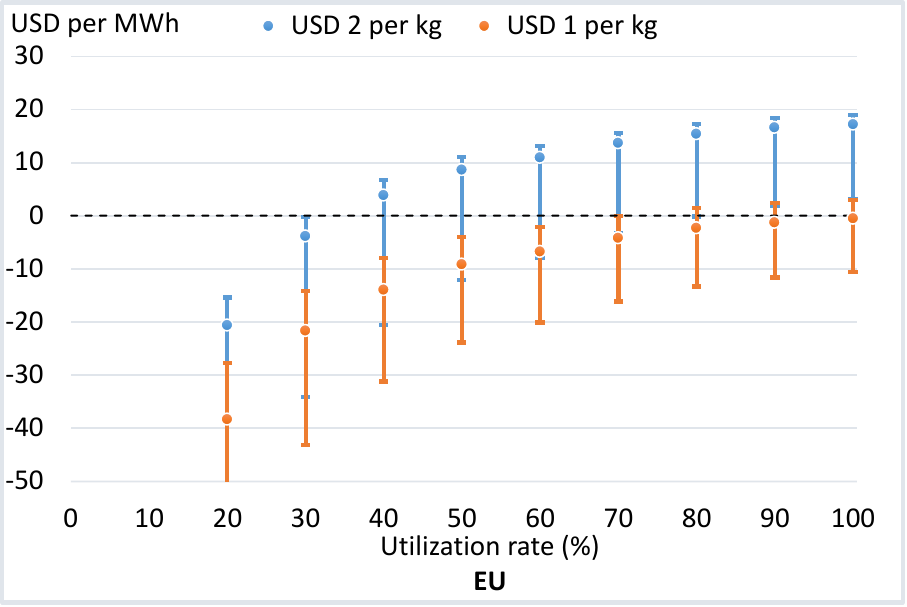}
    \end{subfigure}
    
    %\vspace{0.5cm} % Adjust vertical space between rows of figures
    
    \begin{subfigure}[b]{0.475\textwidth}
        \centering
        \includegraphics[width=\linewidth, height=4cm]{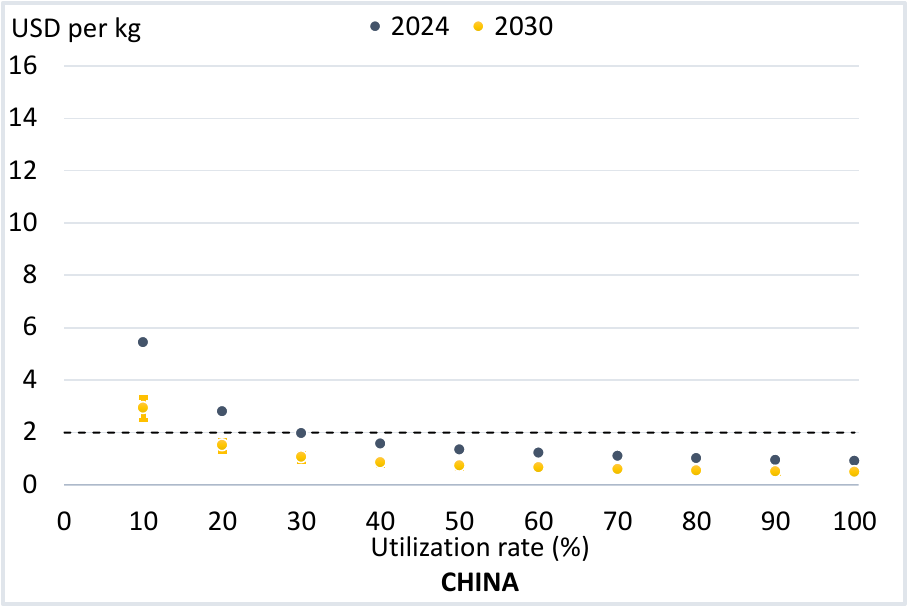}
    \end{subfigure}
    \hfill
    \begin{subfigure}[b]{0.475\textwidth}
        \centering
        \includegraphics[width=\linewidth, height=4cm]{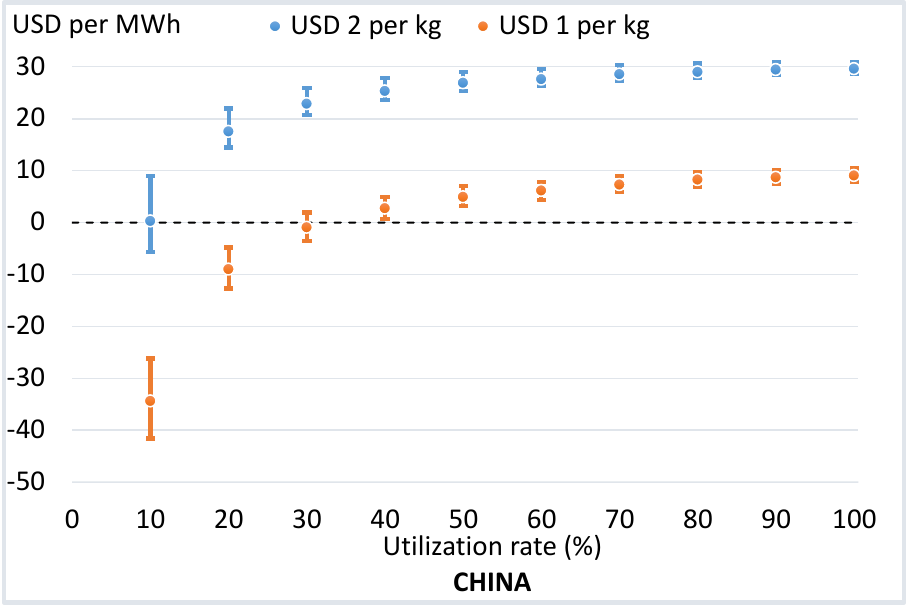}
    \end{subfigure}
    
    %\vspace{0.5cm} % Adjust vertical space between rows of figures
    
    \begin{subfigure}[b]{0.475\textwidth}
        \centering
        \includegraphics[width=\linewidth, height=4cm]{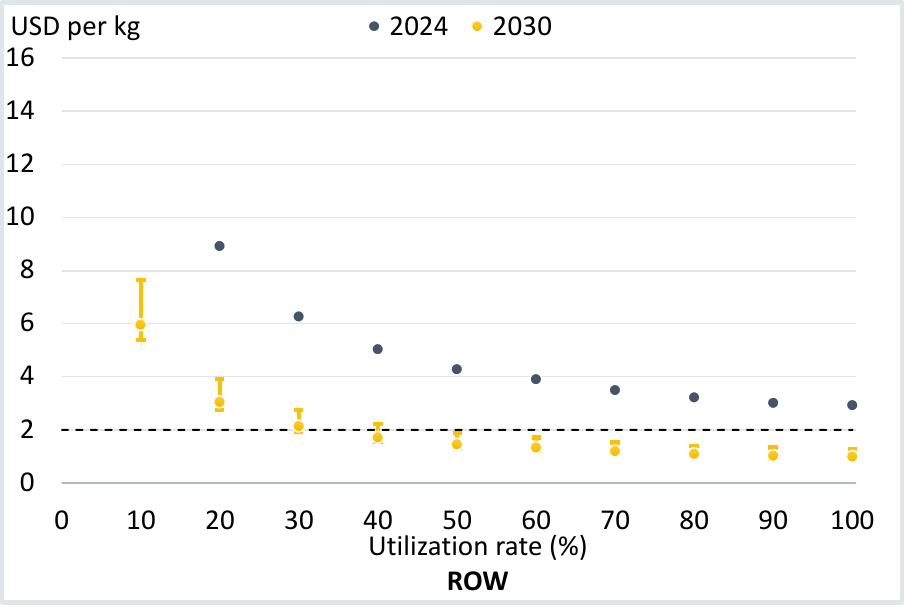}
    \end{subfigure}
    \hfill
    \begin{subfigure}[b]{0.475\textwidth}
        \centering
        \includegraphics[width=\linewidth, height=4cm]{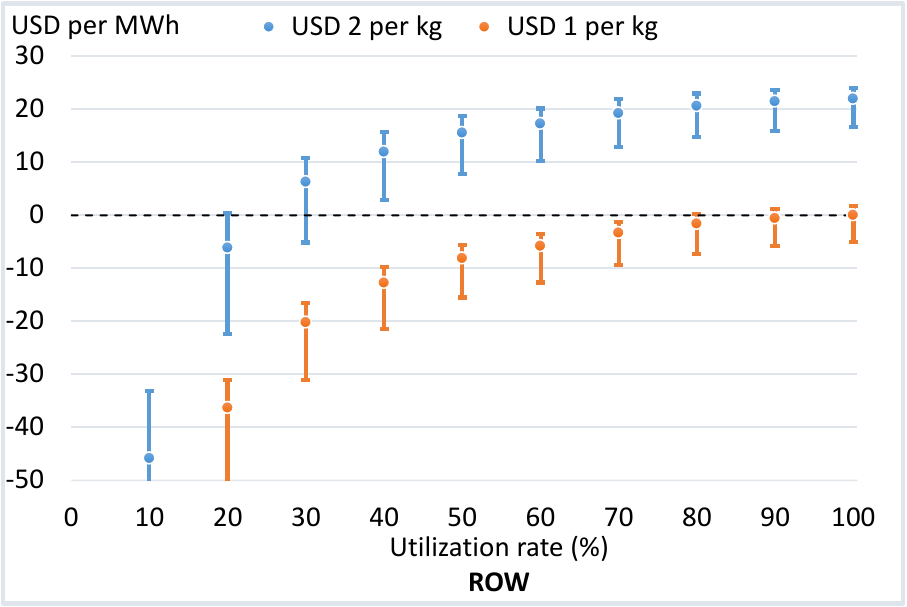}
    \end{subfigure}
    
    \caption{\textbf{Current (dark blue) and projected (yellow) 2030 capital cost contribution to levelized cost of hydrogen by region (left) and required electricity cost to achieve unsubsidized hydrogen production at \$1/kg (orange) or \$2/kg (blue) given 2030 capital cost projections (right).} The error bars depict sensitivity to the projected capital costs of electrolysis.}
    \label{Fig 3:}
\end{figure}

Under stable natural gas and coal prices, the cost of hydrogen production from fossil fuels typically falls between $\$0.5$ and $\$2.5$/kg, depending on the region. The addition of carbon capture and storage increases these costs by up to $\$1$/kg \cite{iea2021hydrogen,iea2022hydrogen}. At present, electrolytic hydrogen production at or below $\$2$/kg is not viable outside of China, as the current contribution of capital and fixed costs to the levelized cost of hydrogen production (LCOH) is near or above $\$2$/kg even at 100\% utilization rates and before factoring in the cost of clean electricity supply ((Fig. \ref{Fig 3:}, left)). Looking forward to 2030, policy-induced capital cost declines ((Fig. \ref{Fig 3:}, left)) could, in principle, enable unsubsidized electrolytic hydrogen production at $\$2$/kg at various combinations of utilization rates and electricity input costs ((Fig. \ref{Fig 3:}, right)). However, reaching a LCOH of $\$1$ to $\$2$/kg, necessary to compete with fossil hydrogen in all regions without ongoing subsidy, appears out of reach in the 2030 time frame as it requires unrealistically low electricity costs. Consequently, while expected cost declines will reduce the magnitude of required policy support, ensuring clean hydrogen from electrolysis displaces fossil-derived hydrogen or is economically attractive in new applications (such as direct reduction of iron for steel-making, synthetic fuels, etc.), will depend on further policy intervention: either ongoing production subsidies, meaningful carbon pricing, or regulations that require decarbonization of industrial and fuel production sectors, particularly in regions with low fossil fuel costs. To illustrate, the 45V hydrogen production subsidy enacted in IRA completely changes the economics of hydrogen production in the US. Combined with learning-by-doing, 45V unlocks cost parity between fossil and electrolytic hydrogen. Indeed, subsidized production costs at or below $\$1-2$/kg are attainable for a range of utilization rates and electricity costs (Fig. \ref{Fig 3:}). For electrolysis projects coupled with dedicated clean energy generation resources, the lower end of utilization rates ($20$ to $40\%$) is achievable with the exclusive use of solar or wind, while the higher end of utilization rates ($60$ to $70\%$) can be achieved with a mix of resources. In addition to 45V, such projects could claim the clean electricity production incentives in IRA, thus ensuring supply of low-cost clean electricity and enhancing the competitiveness of electrolytic hydrogen.\\
Near-term scale-up of water electrolysis may also entail important opportunity costs. Ensuring ample supply of low-cost clean electricity will be critical to the competitiveness of electrolytic hydrogen, but rapidly increasing demand from electrolyzers could hinder the pace of grid decarbonization. Policymakers should therefore balance technology policy goals with respect to electrolysis with the pressing need for grid emission reduction, while considering potential clean electricity supply limitations. This trade-off could be navigated by prioritizing off-grid electrolysis projects in more remote locations where grid interconnection of renewables is challenging and by directing available clean hydrogen to high-impact end-uses.
% \begin{figure*}%
%     \includegraphics[width=0.475\textwidth, height=5cm]{CAPEXContributionUSA.pdf}%
%     \qquad
%     \includegraphics[width=0.475\textwidth, height=5cm]{ElectricityUSA.pdf}%
%     \qquad
%     \includegraphics[width=0.475\textwidth, height=5cm]{CAPEXContributionEU.pdf}%
%     \qquad
%     \includegraphics[width=0.475\textwidth, height=5cm]{ElectricityEU.pdf}%
%     \qquad
%     \includegraphics[width=0.475\textwidth, height=5cm]{CAPEXContributionCHINA.pdf}%
%     \qquad
%     \includegraphics[width=0.475\textwidth, height=5cm]{ElectricityCHINA.pdf}%
%     \qquad
%     \includegraphics[width=0.475\textwidth, height=5cm]{CAPEXContributionROW.pdf}%
%     \qquad
%     \includegraphics[width=0.475\textwidth, height=5cm]{ElectricityROW.pdf}%
%     \caption{\textbf{Current (dark blue) and projected (yellow) 2030 capital cost contribution to levelized cost of hydrogen by region (left) and required electricity cost to achieve unsubsidized hydrogen production at \$1/kg (orange) or \$2/kg (blue) given 2030 capital cost projections (right).} The error bars depict sensitivity to the projected capital costs of electrolysis.}
%     \label{Fig 3:}
% \end{figure*}

\begin{figure*}
    \centering
    \includegraphics[width= \textwidth]{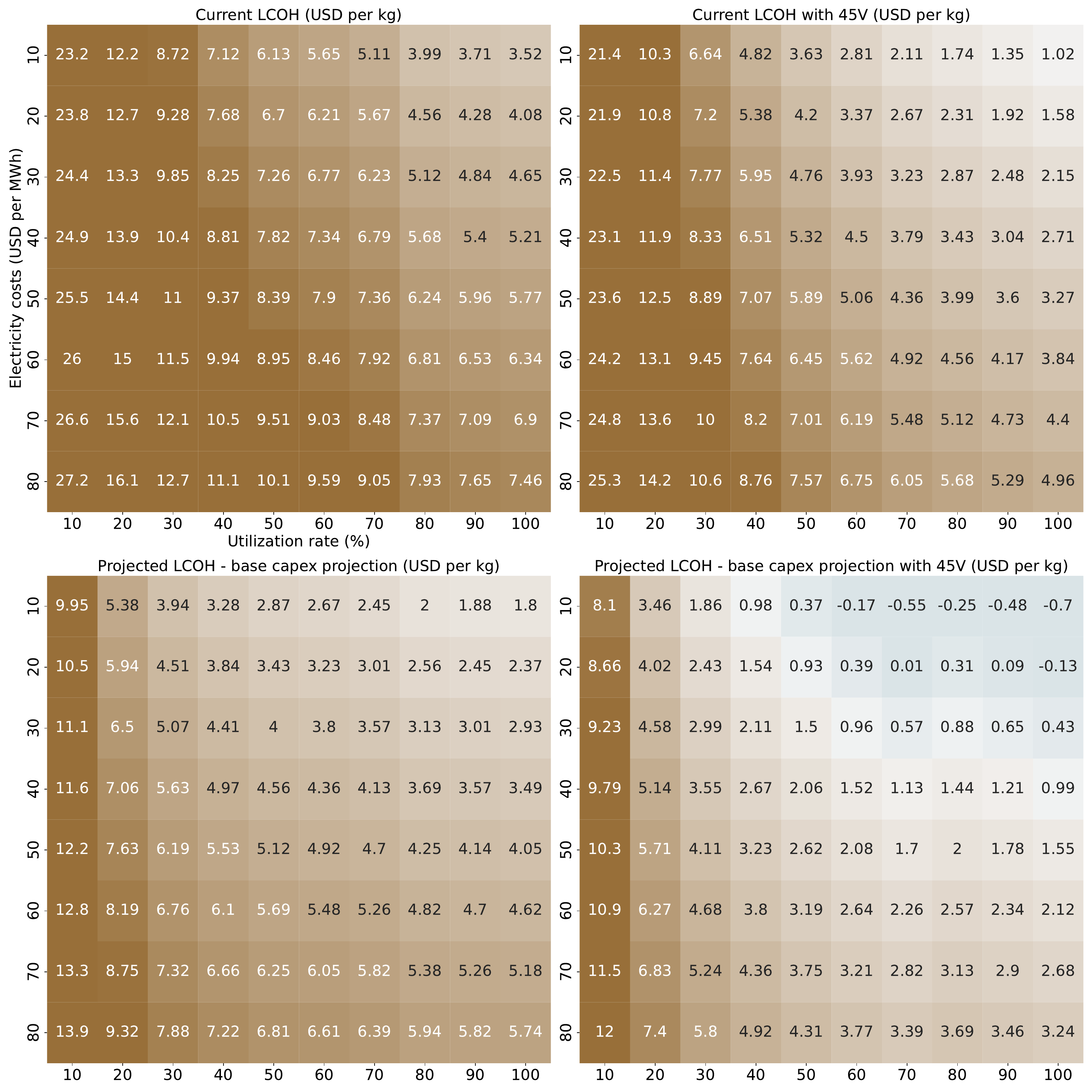}%
    \caption{\textbf{Current and projected LCOH in the United States with and without the 45V production tax credit.}  See supporting information for plots depicting sensitivity to projected electrolysis capital costs.}
    \label{Fig 4:}
\end{figure*}
 
\subsection*{Direct air capture}
Carbon dioxide removal (CDR) is likely necessary to hold long-term warming to 1.5-2$^{\circ}$C by compensating for residual emissions in hard-to-abate sectors and drawing down excess emissions if emission targets are overshot \cite{schiavon-2023}. Direct air capture (DAC) is one class of potentially scalable CDR technologies that capture CO$_2$ directly from the atmosphere using chemical processes. In upcoming years, the industry is expected to transition from the demonstration phase to large-scale deployment as a result of strong policy support from IRA, the Infrastructure and Jobs Act and a growing voluntary procurement market. The first large scale liquid-solvent (L-DAC) facility is currently under construction in Texas, and several other DAC facilities are planned in the US and elsewhere \cite{ieadac}. We focus on L-DAC technology as it is currently scaled up to reach megaton-level capture capacities. We use the global project pipeline (3.5 Mt/y capture capacity by 2030) as an indicator of current policy support and assess the extent to which expected policy-induced deployment can drive L-DAC down an experience curve. While near-term L-DAC deployement is expected to be concentrated in the US, effectively localizing the learning, we employ a learning model whereby cost reductions are a function of the global cumulative deployment capacity, and we use a learning rate empirically observed for sulfur scrubbers (see SI). Our analysis indicates that the successful build-out of the current 2030 project pipeline may result in roughly $12$ to $38\%$ capital cost reductions for L-DAC from \$2600 to \$1600 - 2300 per tCO$_2$/year (Fig. \ref{Fig 1:}). 

\subsubsection*{Cost of removal of carbon dioxide from the atmosphere}
\begin{figure*}%
    \centering
    \includegraphics[width= \textwidth]{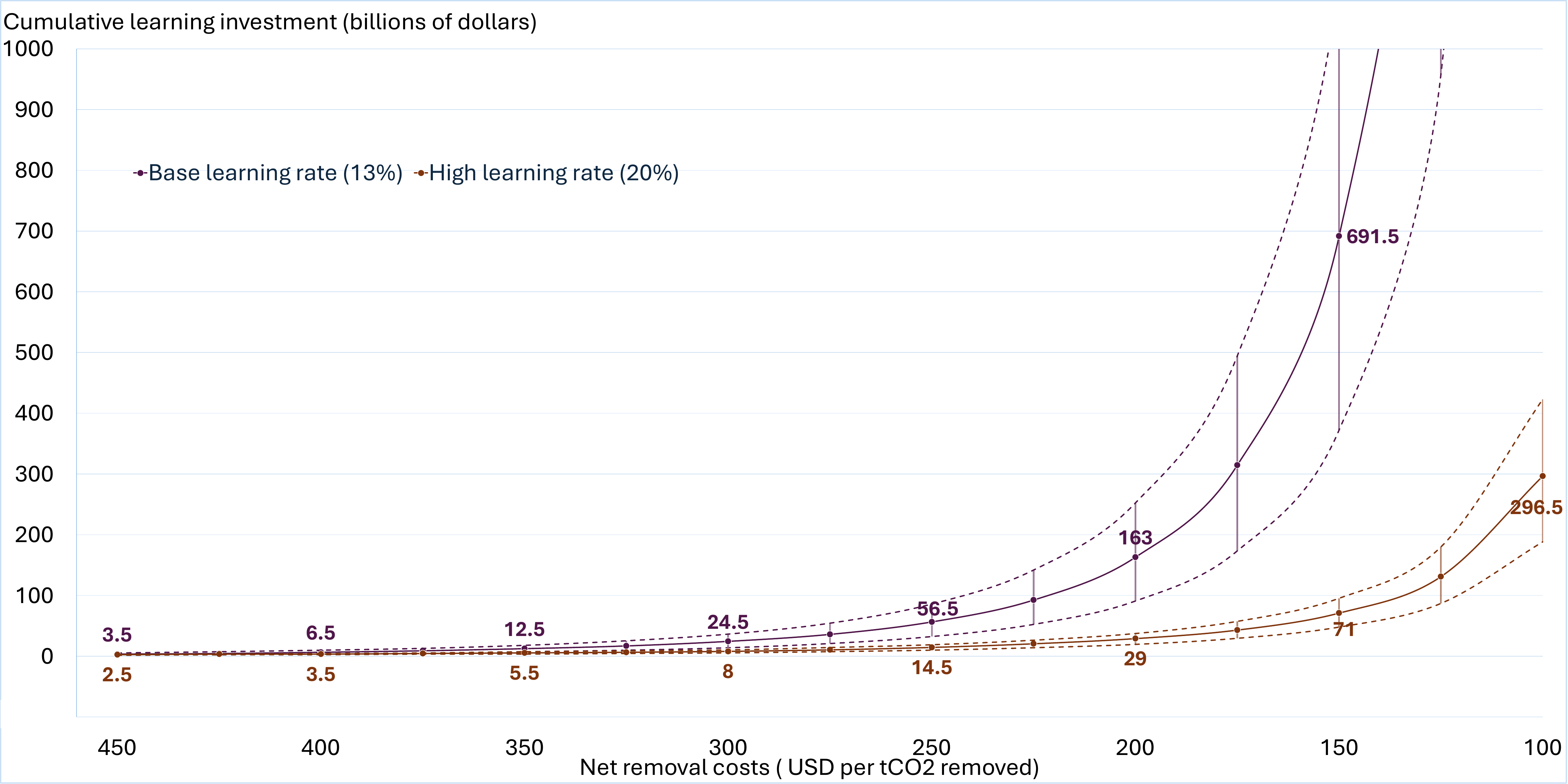}%
    \\
    \includegraphics[width= \textwidth]{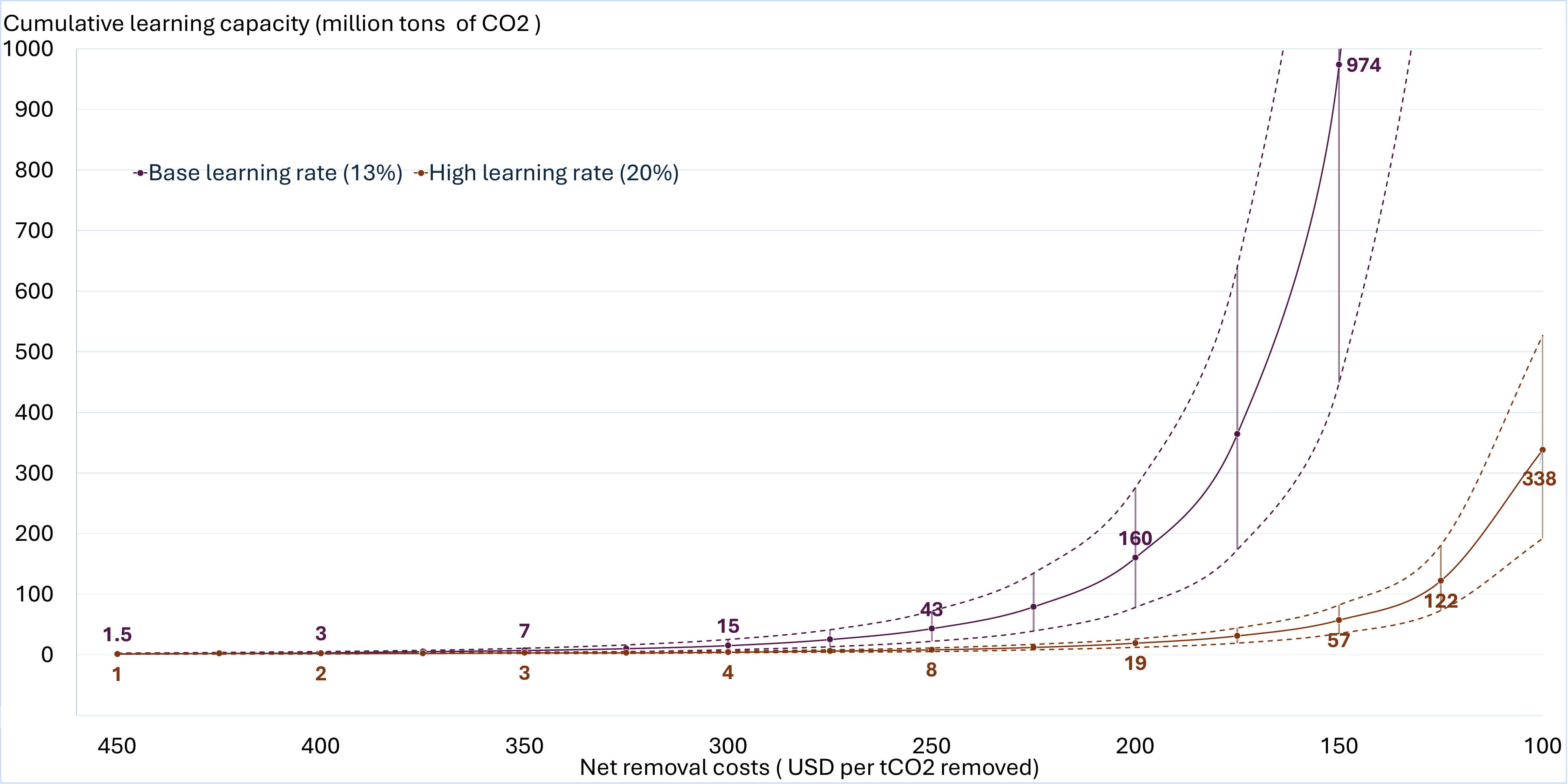}
    \caption{\textbf{Learning investment (top) and capture capacity build-out (bottom) required to reach decreasing net removal cost targets}. The lower and higher dashed curves indicate $0.2\%$ and $3.7\%$ methane leakage rates respectively.}
    \label{Fig 5:}
\end{figure*}

For DAC to play any meaningful role in reaching net-zero targets, the cost of removing carbon dioxide from the atmosphere must be economically viable. L-DAC companies advertise removal costs around $\$400-600$/tCO$_2$ today \cite{oxy2023conference}, and set $\$100/$tCO$_2$ as an industry target \cite{nyt_climeworks}. We seek to assess the feasibility of achieving the net removal cost target and identify the magnitude of the necessary `learning investments' required to drive costs down the experience curve. As planned commercial L-DAC facilities employ natural gas combustion in the capture process, the effect of upstream methane leakage on net removal costs must also be considered.\\
Given the reported costs of the first large-scale L-DAC facility \cite{oxy2023conference}, we estimate current capture costs of roughly $\$483$/tCO$_2$. Depending on upstream methane leakage rates, the net removal costs, however, can range anywhere between the aforementioned cost, to as much as $\$610$/tCO$_2$ for leakage rates up to $3.7\%$ under GWP100. By 2030, we project net removal costs to range between $\$346$ and $\$427$/tCO$_2$. The estimation of the learning investment reveals that $\$100/$tCO$_2$ target is likely unattainable for L-DAC. Indeed, the investment and capacity build-out necessary to reach the aspirational industry target are unrealistically large, even under low upstream methane leakage rates, as illustrated in Fig. \ref{Fig 5:}. Learning effects are stronger initially when cumulative capacity doubling is easier to achieve. As more capacity is built, the resulting marginal reduction in cost per unit of capacity deployed naturally decreases, requiring larger sums of investment to achieve the next increment of cost reduction. For the $\$100$/tCO$_2$ removal cost to be achievable, L-DAC would have to follow a high learning rate experience curve ($\sim20\%$), typically observed only for mass-manufactured, modular technologies as opposed to large-scale complex industrial facilities. Even in this case, the level of required cumulative investment is substantial (on the order of \$300 billion cumulatively to deploy roughly 338 million metric tonnes/year of capture capacity under base case assumptions).
\newline
Fig. \ref{Fig 5:} also underscores the paramount importance of tackling fugitive emissions. From an economic point of view, minimizing leakage rates ensures that the net removal costs align closely with capture costs, therefore significantly reducing the learning investment needed to diffuse the technology. Consequently, a net removal cost of $\$200$/tCO$_2$ would eventually be within reach, though still necessitating substantial investments (on the order of \$163 billion cumulatively to deploy roughly 160 million metric tonnes/year of capture capacity under base case assumptions). However, addressing methane leakage goes beyond just economics, as it is likely to determine the technology's very legitimacy. With a recent wave of criticism and mounting evidence against ineffective carbon offset programs \cite{guardian2023carbonoffsets}, DAC aims to establish itself as a reliable and verifiable provider of carbon offsets, but significant upstream emissions cast doubt on the extent to which L-DAC can live up to that. A recent study estimated methane leakage levels of $9\%$ in the New Mexico Permian Basin \cite{chen_quantifying_2022}. For context, if one considers the short-term warming potential of methane (e.g. GWP20), the emissions resulting from a $7\%$ methane leakage rate would be enough to cancel out \textit{all} of the CO$_2$ captured from a natural gas-fueled L-DAC facility. This underscores the critical need for regulation to detect and account for leaks in the gas supply chain as well as rewarding government subsidies or voluntary purchases based on \textit{net} CO$_2$ removed, inclusive of fuel cycle emissions, rather than gross CO$_2$ captured. Such an approach would incentivize DAC developers to minimize upstream emissions whenever possible, including reductions of methane leakage for natural gas-fueled L-DAC facilities or use of clean energy inputs in lieu of natural gas. This could be achieved by using renewable energy coupled with resistance heating and high-temperature thermal storage to run the air capture process instead of burning natural gas, or by purchasing carbon-free electricity while ensuring strict adherence to temporal and geographical matching and additionality requirements \cite{ricks_minimizing_2023}. Solid sorbent DAC systems require low-grade heat (around 100$^{\circ}$C) for regeneration and could thus also be supplied with heat from co-located geothermal or nuclear energy or sources of waste heat \cite{mcqueen-2021}.\\
\subsection*{Synthetic kerosene}
\begin{figure*}[htbp]
    \centering
    
    \begin{subfigure}[b]{0.475\textwidth}
        \centering
        \includegraphics[width=\linewidth]{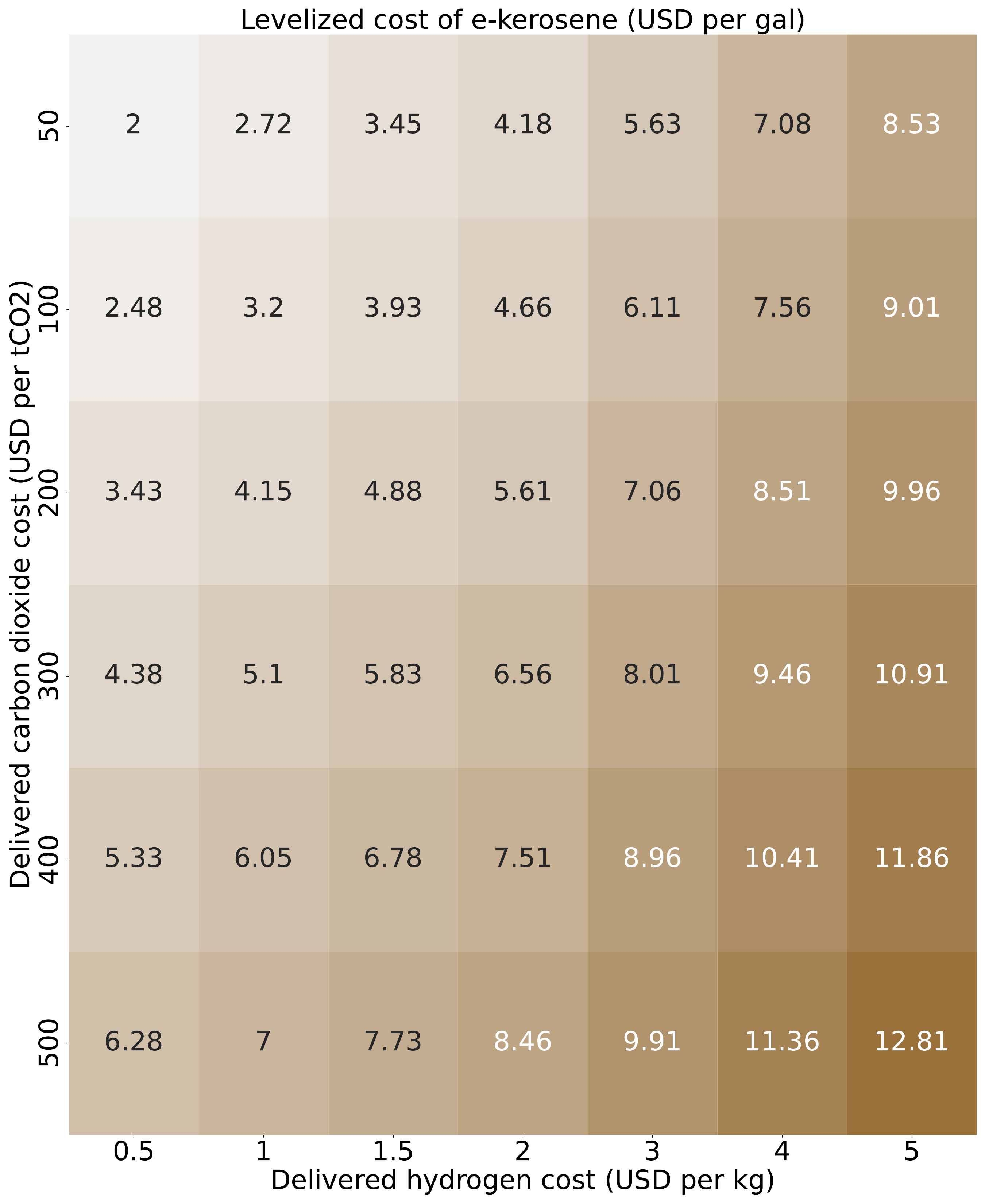}
        \caption{Levelized cost of e-kerosene production}
        \label{subfig:lco_ek}
    \end{subfigure}
    \hfill
    \begin{subfigure}[b]{0.475\textwidth}
        \centering
        \includegraphics[width=\linewidth]{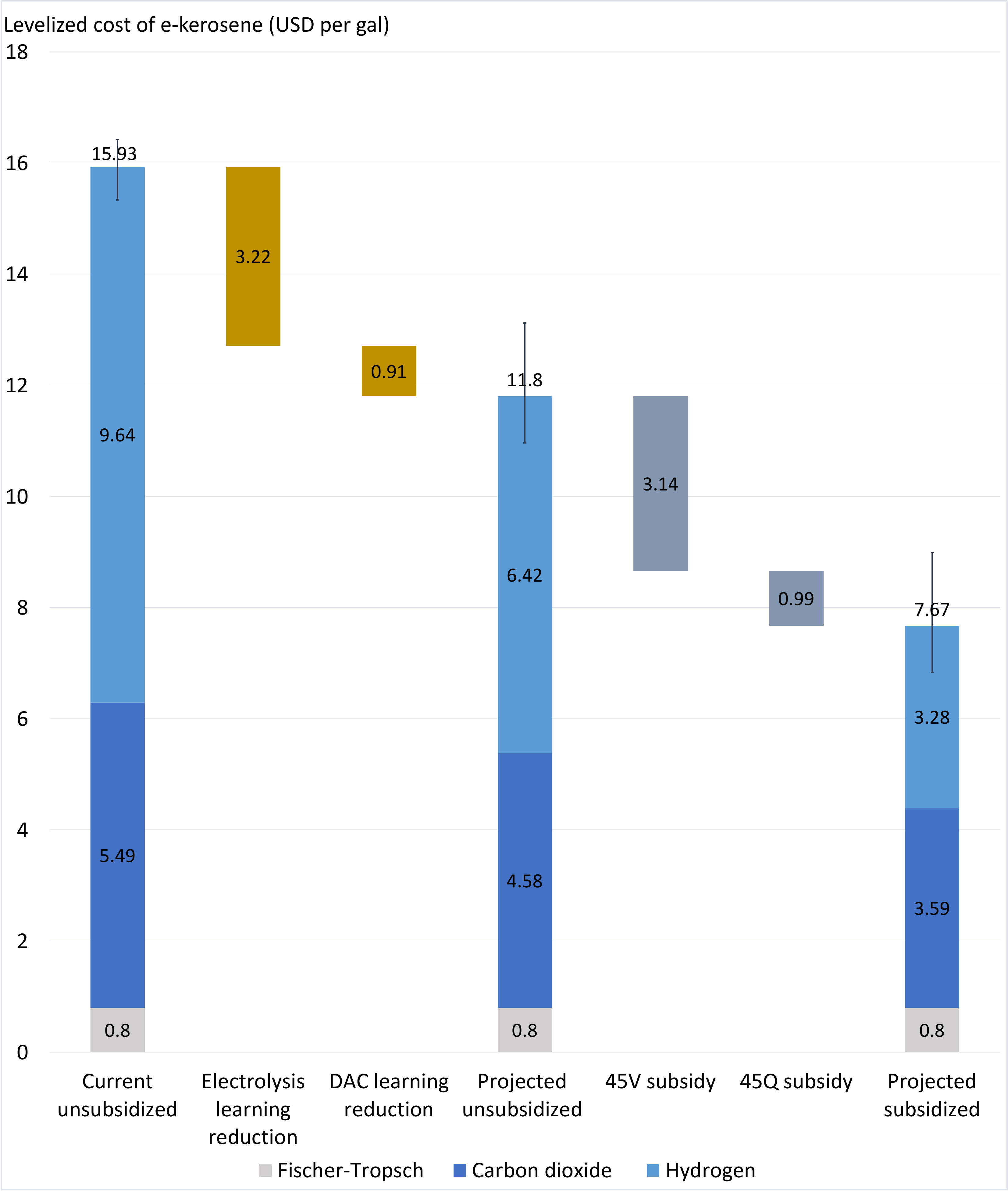}
        \caption{Effect of learning and IRA subsidies on e-kerosene costs}
        \label{subfig:electrofuel}
    \end{subfigure}
    
    \caption{\textbf{The levelized cost of e-kerosene production as a function of carbon dioxide and hydrogen costs (left). The combined effect of learning and IRA subsidies on e-kerosene costs in the US (right)}. The assumed electricity costs are \$50/MWh. The error bars indicate sensitivity to L-DAC and electrolysis costs.}
    \label{Fig 6:}
\end{figure*}

The aviation industry is responsible for about $2\%$ of global greenhouse gas emissions \cite{ritchie_climate_2023} and aviation demand is growing rapidly \cite{icao2022envreport}. In recent years, blending fossil jet fuel with low-carbon `sustainable aviation fuels' (SAF) produced from a range of different pathways has emerged as one of the main decarbonization pillars for this sector. In the near-term, the bulk of SAF supply will come from used oil and fats \cite{icf_sustainable_aviation_fuel}. In the long run, the share of second-generation SAF from bio-feedstocks such us woody biomass and municipal solid waste is expected to increase. However, supplies of feedstock for these biofuel production pathways are limited and several sectors including aviation and heat and power generation compete for them \cite{icf_sustainable_aviation_fuel}. The only fuel pathway without inherent feedstock constraints would involve e-kerosene produced from carbon-free hydrogen and CO$_2$ from DAC. We thus investigate how the anticipated cost declines in electrolysis and DAC discussed above will impact e-kerosene production costs.\\
At present, we estimate the levelized cost of producing e-kerosene (LCOek) to be between $\$11$ and $\$16$/gal ($\$2.9$ and $\$4.2$/l) depending on the region (see SI). By 2030, anticipated cost reductions for electrolysis and L-DAC would only reduce the LCOek by $\$1$ to $\$4$/gal ($\$0.26$ to $\$1$/l), leaving synthetic kerosene several times more expensive than fossil kerosene (which has ranged from \$1-3.25/gal in the US from 2013 to 2018 \cite{eiajetfuel}). The production of synthetic kerosene at competitive costs requires a combination of low-cost hydrogen and low carbon removal costs ((\ref{Fig 6:}, left)), neither of which are attainable in 2030 and potentially beyond without some form of ongoing policy support. Indeed, government subsidies such as incentives in IRA can substantially reduce the cost premium of e-kerosene ((\ref{Fig 6:}, right)).\\
However, at very low blending levels envisioned for the 2030 decade, the higher cost of unsubsidized synthetic kerosene is unlikely to significantly affect aviation costs. At projected e-kerosene costs and a $5\%$ blending level consistent with the ReFuelEU Aviation mandate for 2035, we estimate that the fuel costs of a flight connecting New York to London would increase by $\$14$ to $\$30$ per passenger depending on the cost of fossil jet fuel and the efficiency of the aircraft (see SI). For reference, the average fare of such a flight varies between roughly $\$160$ and $\$320$ \cite{whereandwhen_flights_london}. The impact of e-kerosene on flight costs becomes pronounced at higher blending levels, particularly for inefficient aircraft that burn more fuel per passenger-kilometer (see SI).\\ %Increasing blending mandates could potentially stimulate more demand for higher efficiency aircraft.\\
Synthetic kerosene could alternatively be produced using a combination of electrolysis and CO$_2$ captured from industrial sources such as cement or ethanol production, thus benefiting from lower carbon dioxide costs. However, unless sourced from biogenic CO$_2$, this pathway is not a viable option for deep decarbonization and would require stringent carbon accounting standards to limit double counting of emissions reductions by both industrial source and SAF producer.\\
Beyond economics, the production of synthetic fuels from a combination of DAC and electrolysis requires large amounts of energy and infrastructure. Despite no fundamental constraints on feedstock availability, whether synthetic fuel production can be scaled up in practice to meet a meaningful portion of global jet fuel demand is an open question, as this would require hundreds of gigawatts of electrolysis and low-carbon electricity and hundreds of millions of tons/yr of CO$_2$. The scale of resources required, and the significant costs, prompts questions about the extent to which e-kerosene can be relied upon in decarbonization strategies. This underscores the need to consider policies that also encourage alternative decarbonization strategies for the aviation sector, including electrification of short-haul flights and/or demand-side measures such as the promotion of alternative modes of transport when possible. 
% \begin{figure*}%
%     \includegraphics[width=0.475\textwidth]{LCOek.pdf}
%     \qquad
%     \includegraphics[width=0.475\textwidth]{Electrofuel.pdf}
%     \caption{\textbf{The levelized cost of e-kerosene production as a function of carbon dioxide and hydrogen costs (left). The combined effect of learning and IRA subsidies on e-kerosene costs in the US (right)}. The assumed electricity costs are $\$50$/MWh. The error bars indicate sensitivity to L-DAC and electrolysis costs.}
%     \label{Fig 6:}
% \end{figure*}

\section*{Conclusion}
In this article, we explored how the current policy environment could drive cost reductions of electrolysis, DAC, and e-kerosene, three nascent technologies that may play an important role in long-term deep decarbonization and climate mitigation. Water electrolysis cost are likely to fall substantially by 2030, though electrolytic hydrogen is unlikely to be competitive with fossil hydrogen in this time frame without ongoing policy support. The cost of electrolytic hydrogen will vary regionally, dictated by local factors including the abundance of low-cost clean energy, the cost of fossil fuels, local regulation, and the cumulative experience of local workforces and supply chains. There appears to be no viable pathway for liquid-solvent based DAC to reach a CO$_2$ removal cost target of \$100/tCO$_2$, though \$200/tCO$_2$ may eventually be within reach. DAC's pathway to viable removal costs will require substantial cumulative investments, and there is an urgent need to establish carbon accounting standards that account for fuel cycle emissions and their impact on net removal costs, such as methane leakage for natural gas-fueled liquid DAC facilities. As for e-kerosene, costs will remain many times higher than conventional kerosene, despite potential cost reductions in input hydrogen from electrolysis and CO$_2$ from DAC. The high cost of the fuel and the scale of input resources required to produce a meaningful share of global aviation fuel demand cast doubt on the potential for e-kerosene to offer a viable route to decarbonize aviation. It is also important to remember that experience-related cost reductions calculated in this study are not a certain outcome. The challenge of scaling emerging technologies and unlocking learning effects remains a formidable task requiring sustained support under a stable policy environment, coordination across different stakeholders and sectors, proactive planning, and overcoming obstacles related to permitting and securing project finance.
\newpage
\section*{Experimental Procedures
}
\subsection*{Experience curve model for water electrolysis}
As discussed in the main manuscript, the capital costs of a water electrolysis project consists of three parts: the electrolyzer stack, the
balance of plant (BoP), and engineering, procurement, and construction (EPC).\\
The stack costs are assumed to follow a global experience curve formulated mathematically as:
\begin{equation*}
    C_{tech,projected}^{global} = C_{tech,current}^{global} \times \left(\frac{x_{tech,projected}^{global}}{x_{tech,current}^{global}}\right)^{\frac{\log\left(1-\alpha_{tech}^{global}\right)}{\log(2)}}
\end{equation*}
where $tech$ stands for Western PEM, Chinese PEM, Western ALK or Chinese ALK, $\mathrm{C}_{\text{tech,current}}^{\text{global}}$ and $\mathrm{C}_{\text{tech,projected}}^{\text{global}}$ are the current and projected costs of stack type $tech$, $\mathrm{x}_{\text{tech,current}}^{\text{global}}$ and $\mathrm{x}_{\text{tech,current}}^{\text{global}}$ are the current and projected global cumulative deployed capacities of stack type $tech$, and $\alpha_{\text{tech}}^{\text{global}}$ is the global learning rate of stack type $tech$.\\
The BoP and PEC costs are assumed to follow a local experience curve formulated mathematically as:
\begin{equation*}
     C_{BOP\&PEC,projected}^{reg,tech} =  C_{BOP\&PEC,current}^{reg,tech} \times \left(\frac{x_{projected}^{reg}}{x_{current}^{reg}}\right)^{\frac{\log\left(1-\alpha_{BOP\&PEC}^{reg}\right)}{\log(2)}}
\end{equation*}
where $reg$ stands for USA, CHINA, EU or ROW, $\mathrm{C}_{\text{BOP\&PEC,current}}^{\text{reg,tech}}$ and $\mathrm{C}_{\text{BOP\&PEC,projected}}^{\text{reg,tech}}$ are current and projected BoP and PEC capital costs, $\mathrm{x}_{\text{projected}}^{\text{reg}}$ and $\mathrm{x}_{\text{projected}}^{\text{reg}}$ are current and projected cumulative deployed capacity of all electrolyzer stack types in region $reg$, and $\alpha_{\text{BOP\&PEC}}^{\text{reg}}$ is the local BOP and PEC learning rate of region $reg$.\\ 
Finally, the projected capital costs of a water electrolysis project $\mathrm{C}_{\text{tech,projected}}^{\text{reg}}$ in region $reg$ using stack type $tech$ are given by:
\begin{align*}
    C_{tech,projected}^{reg} &=   C_{tech,projected}^{global} + C_{BOP\&PEC,projected}^{reg} \\ 
                                     &=  C_{tech,current}^{global} \times \left(\frac{x_{tech,projected}^{global}}{x_{tech,current}^{global}}\right)^{\frac{\log\left(1-\alpha_{tech}^{global}\right)}{\log(2)}}  \\
                                     &+ C_{BOP\&PEC,current}^{reg,tech} \times \left(\frac{x_{projected}^{reg}}{x_{current}^{reg}}\right)^{\frac{\log\left(1-\alpha_{BOP\&PEC}^{reg}\right)}{\log(2)}}
\end{align*}
\subsection*{Experience curve model for direct air capture}
The liquid-solvent direct air capture technology (L-DAC) is assumed to follow a global learning curve, mathematically represented as:
\begin{equation*}
    C_{projected}^{L-DAC} = C_{current}^{L-DAC} \times \left(\frac{x_{projected}^{L-DAC}}{x_{current}^{L-DAC}}\right)^{\frac{\log\left(1-\alpha_{L-DAC}\right)}{\log(2)}}
\end{equation*}
where $\mathrm{C}_{\text{current}}^{\text{L-DAC}}$ and $\mathrm{C}_{\text{projected}}^{\text{L-DAC}}$ are the current and projected capital costs of an L-DAC facility, $\mathrm{x}_{\text{current}}^{\text{L-DAC}}$ and $\mathrm{x}_{\text{projected}}^{\text{L-DAC}}$ are current and projected global cumulative installed L-DAC capacity, and $\alpha_{\text{L-DAC}}$ is the global L-DAC learning rate.\\
Additional details on modeling assumptions can be found in the supplementary information.
\bibliography{Citations}% common bib file
\end{document}